\documentclass{article}
\usepackage{amssymb}
\usepackage{amsmath}
\usepackage{amsthm}
\usepackage{amsfonts}
\usepackage{graphicx}
\usepackage{geometry}

\newtheorem{theorem}{Theorem}

\newtheorem{remark}[theorem]{Remark}

\renewenvironment{proof}[1][Proof]{\noindent\textbf{#1.} }{\
\rule{0.5em}{0.5em}}
\numberwithin{equation}{section}
\input{tcilatex}
\begin{document}

\title{How to make Dupire's local volatility work with jumps}
\author{Peter K. Friz\thanks{%
Corresponding author: friz@math.tu-berlin.de, friz@wias-berlin.de}, Stefan
Gerhold and Marc Yor \\
TU and WIAS Berlin, TU Wien, Universit\'e Pierre-et-Marie-Curie (Paris VI)}
\maketitle

\begin{abstract}
There are several (mathematical) reasons why Dupire's formula fails in the
non-diffusion setting. And yet, in practice, ad-hoc preconditioning of the
option data works reasonably well. In this note we attempt to explain why.
In particular, we propose a regularization procedure of the option data so
that Dupire's local vol diffusion process recreates the correct option
prices, even in manifest presence of jumps.
\end{abstract}

\section{Failure of Dupire's formula in non-diffusion setting}

\label{se:intro}

Local volatility models \cite{Du94, Du96}, ``$dS/S=\sigma _{\mathrm{loc}%
}( S,t) dW$'', are a must-be for every equity option trading floor.
A central ingredient is \emph{Dupire's formula}, which allows to obtain the
diffusion coefficient $\sigma _{\mathrm{loc}}$ directly from the market (or
a more complicated reference model),%
\begin{equation}
\sigma _{\mathrm{loc}}^{2}(K,T)=2\partial _{T}C/K^{2}\partial _{KK}C,
\label{DupForm}
\end{equation}%
in terms of\ (call) option prices at various strikes and maturities.\footnote{Throughout, we work
under the appropriate forward measure to avoid drift terms.}
Of course, there are ill-posedness issues how to compute derivatives when only
given discrete (market) data; this inverse problem is
usually solved by fitting market (option or implied vol) data via a smooth
parametrization, from which $\sigma _{\mathrm{loc}}$ is then computed.

On a more fundamental level, given an arbitrage-free option price surface%
\begin{equation*}
\left\{ C( K,T) :K\geq 0,T\geq 0\right\},
\end{equation*}%
there are two problems in a non-diffusion setting:

\begin{itemize}
\item \emph{Lack of smoothness} ($\partial _{KK}C$ is precisely the stock price
density). For instance, the asymmetric Variance Gamma (AVG; cf.~\cite{CoVo05}%
) process $\left( X_{t}\right) $ has characteristic function%
\begin{align*}
\mathbb{E}[ \exp ( iuX_{T}) ]  &=\left( \frac{1}{%
1-i\theta \nu u+\left( \sigma ^{2}\nu /2\right) u^{2}}\right) ^{T/\nu } \\
&\sim const\cdot u^{-2T/\nu }\quad\text{ as }\ u\rightarrow \infty.
\end{align*}%
When $2T/\nu >1$ (equivalently: $T>\nu /2$), the characteristic function is
integrable on $\mathbb{R}$ and $X_{T}$ admits a continuous density; the
same is then true for $S_{T}=S_{0}\exp \left( X_{T}+\mu T\right) $. This is not
the case, however, for $T<\nu /2$, and indeed the density (given explicitly
e.g.\ in~\cite[p.~82]{MaCaCh98}) has a singularity
at the origin, so that $C$ cannot possibly be twice continuously
differentiable in $K$.

\item \emph{Blowup of the short end of the local volatility surface.} Assume, for
the sake of argument, that%
\begin{equation*}
S_{T}=S_{0}\exp ( X_{T}+ \sigma W_{T}+\mu T),
\end{equation*}%
where $X$ is a pure jump L\'{e}vy process with L\'{e}vy measure $L=L(dx)$,
and~$W$ is a standard Brownian motion. For instance, in the Merton jump diffusion model,
where $X_{T}$ has normally distributed jumps, one can see 
\begin{equation}\label{eq:merton blowup}
\sigma _{\mathrm{loc}}^{2}(K,T)\approx const \cdot \frac{1}{T}\quad \text{ as }\ T\to
0,\,K\neq S_{0}.
\end{equation}%
In fact, this blowup can be fully quantified, thanks to a recently established saddle point
formula \cite{DeFrGe13}; details are given in the appendix.
There we show that the approximation~\eqref{eq:merton blowup} is also true in the
NIG (normal inverse Gaussian) model, whereas in Kou's double exponential
jump diffusion model the blowup is of order $T^{-1/2}$.
 (The analysis
is not restricted to exponential L\'{e}vy models. Indeed, general jump diffusion
resp.\ semimartingales are ``tangent'', at a given point in space-time, to a L\'{e}vy
process, cf.~\cite{JaSh03,St75}.)

 \end{itemize}

At least from a mathematical point of view, all this is strong evidence that (%
\ref{DupForm}) is not meaningful in presence of jumps. In fact, if one views
(\ref{DupForm}) as a (forward) PDE for call option prices as function of $K,T$,
then the analogous formula in a jump setting is a (forward)\ PIDE, which
is thus the ``natural generalization'' of Dupire's formula in presence of
jumps; cf.~\cite{BeCo10} and the references therein, in particular towards
so-called local L\'{e}vy models \cite{CaGeMaYo04}.

For good or bad, practitioners use (\ref{DupForm}) no matter what. Smoothness
is usually not an issue since local vol is typically obtained from a
smooth parametrization of market (implied vol) data.\footnote{%
Doing so without introducing arbitrage is decisively non-trivial, see~\cite%
{GaJa11,GaJa12}.} On the other hand, the short-time blowup of
local volatility,
$ \sigma _{\mathrm{loc}}(\cdot,T) \to \infty$ as $T \to 0$,
 is an immediate obstacle, for it already makes it unclear to
what extent there exists a (unique) strong solution to the stochastic differential equation
$dS/S=\sigma _{\mathrm{loc}}\left( S,t\right) dW$, let alone how to sample from it.
 It is reported from practitioners that various ad-hoc truncation and
mollification procedures are in place, say with parameter $\varepsilon $,
after which the local volatility is not explosive anymore; Monte Carlo
simulations of ``$dS^{\varepsilon }/S^{\varepsilon }=\sigma _{\mathrm{loc}%
}^{\varepsilon }( S^{\varepsilon },t) dW$'' are then possible.
There is every practical evidence that%
\begin{equation*}
\mathbb{E}[( S_{T}^{\varepsilon }-K) ^{+}]\approx C( K,T)
\end{equation*}%
is a good approximation (for otherwise, risk management would not allow
this in practice). And yet, to our knowledge, there has been no mathematical
justification to date for this type of approximations.

\section{Regularization of call prices: how to make Dupire work}

We now consider the situation of a given martingale $( S_{t}) $
which creates a smooth call price surface, with (strict) absence of calendar
and butterfly spreads.\ This situation is typical in the industry. (Of
course, only option data is known, but upon a suitable parametrization
thereof, one is in the just described situation.) In particular, $\sigma _{%
\mathrm{loc}}^{2}(K,T)=2\partial _{T}C/K^{2}\partial _{KK}C$ is
well-defined, as long as $T>0$, but may explode as $T\to 0$ (thereby
indicating the possibility of jumps). This situation is also typical in a
generic non-degenerate jump diffusion setting.\footnote{%
The mathematics here is well understood; a smooth density (and then call
prices) can be the result of a (hypo)elliptic diffusion part, infinite
activity jumps may also help.
From a practical point of view, models tend to be locally elliptic with jumps super-imposed, so that we shall not pursue
further technical conditions here.}

\bigskip

\begin{theorem}
Assume that $\left( S_{t}\right) $ is a martingale (possibly with jumps) with
associated smooth call price surface $C$,%
\begin{equation*}
\forall K,T\geq 0:C(K,T) =\mathbb{E}[\left( S_{T}-K\right) ^{+}],
\end{equation*}%
such that $\partial _{T}C>0$ and $\partial _{KK}C>0$, i.e.\ (strict) absence
of calendar and butterfly spreads. Define $\varepsilon $-shifted local
volatility
\begin{equation*}
\sigma _{\varepsilon }^{2}( K,T) =\frac{2\partial _{T}C(K,
T+\varepsilon) }{K^{2}\partial _{KK}C(K, T+\varepsilon ) 
}.
\end{equation*}%
Then $dS^{\varepsilon }/S^{\varepsilon }=\sigma _{\varepsilon}
( S^{\varepsilon },t) dW$, started at randomized spot $%
S_{0}^{\varepsilon }$ with distribution%
\begin{equation*}
\mathbb{P}[ S_{0}^{\varepsilon }\in dK] /dK=\partial _{KK}C(K,
\varepsilon),
\end{equation*}%
admits a unique, non-explosive strong SDE solution such that%
\begin{equation*}
\forall K,T\geq 0:\mathbb{E}[( S_{T}^{\varepsilon }-K) ^{+}]\rightarrow
C(K,T)\quad \text{ as }\ \varepsilon \rightarrow 0\text{.}
\end{equation*}
\end{theorem}

Our assumptions encode that the model itself, i.e.\ the specification of the
dynamics of $S$, has regularization effects built in. The result is then, in
essence, a variation of the arguments put forward in a recent revisit of
Kellerer's theorem, see~\cite{HiRoYo12} and the references therein.

\begin{proof}
By assumption, $\sigma _{\varepsilon }^{2}$ is well-defined for all $%
T\geq 0$%
. (In general, i.e.\ without adding $\varepsilon $ and in presence
of jumps, local
vol is not well-defined in the sense that it may blow up as 
$T\rightarrow 0$%
.) Existence of a unique non-explosive local vol SDE
solution for continuous
and locally bounded diffusion coefficient is a
classical result of the
theory of (one-dimensional) SDEs. Set $%
a^{\varepsilon }( S, t ) =%
\frac{1}{2}S^{2}\sigma _{\varepsilon
}^{2}( S, t) ;$ the generator
of $S^{\varepsilon }$ reads $%
L=L_{t}=a^{\varepsilon }\partial _{SS}$. Set
also $C^{\varepsilon }(K,T) = C(K, T+\varepsilon ) $
and $p^{\varepsilon }=\partial
_{KK}C^{\varepsilon },$ the density of $%
S_{T+\varepsilon }$. By definition,%
\begin{equation*}
a^{\varepsilon
}( K,T) =\frac{\partial _{T}C^{\varepsilon }(K,T) }{%
p^{\varepsilon }( K,T) },
\end{equation*}%
and hence, using $%
\partial _{TKK}C^{\varepsilon }( K,T)
=\partial
_{T}p^{\varepsilon }$,%
\begin{equation}\label{eq:fokker}
\partial _{KK}\left(
a^{\varepsilon }p^{\varepsilon }\right) =\partial
_{T}p^{\varepsilon }.
\end{equation}%
In particular, $p^{\varepsilon}$ (which is $L^{1}\left( 0,\infty
\right) \cap
C^{\infty }\left( 0,\infty \right)$  in $K$) is a (classical) solution
to the above
Fokker-Planck equation. On the other hand, $S^{\varepsilon }$
solves the
martingale problem for $L_{t}$ in the sense that for any test
function $%
\varphi $, say, smooth with compact support,
\begin{equation*}%
t \mapsto \varphi \left( S_{t}^{\varepsilon }\right) -\varphi
\left(
S_{0}^{\varepsilon }\right) -\int_{0}^{t}L_{s}\varphi
\left(
S_{s}^{\varepsilon }\right) ds
\end{equation*}%
is a zero-mean
martingale. Taking expectations and writing $q^{\varepsilon
}=q^{\varepsilon }(dS,t) $ for the law of $S_t^{\varepsilon }$,
noting $q^{\varepsilon }( dS,0) =p^{\varepsilon }(S, 0)dS$,
we see 
\begin{equation*}
\int \varphi ( S) q^{\varepsilon}( dS,t) =\int
\varphi ( S) q^{\varepsilon }(
dS,0) +\int_{0}^{t}\int
a^\varepsilon ( S,s) \varphi ^{\prime \prime
}( S) q^{\varepsilon}( dS,s),
\end{equation*}
which is nothing but an (analytically weak, in space) formulation of
the
 Fokker-Planck equation~\eqref{eq:fokker}, with solution given in terms of a family
of
measures, $q^{\varepsilon }=\{q^{\varepsilon }( dS,t) :t\geq
0\}$%
. By a suitable uniqueness result for such equations due to M. Pierre
(see~\cite{HiRoYo12}), we see that $p^{\varepsilon }dS=q^{\varepsilon }$. And
we have
the following consequence for call prices based on $\varepsilon $%
-regularized local vol:
\begin{align*}
\mathbb{E}[\left( S_{T}^{\varepsilon
}-K\right) ^{+}] &=\int \left( S-K\right)
^{+}q^{\varepsilon }(dS,T)  \\
&=\int \left( S-K\right) ^{+}p^{\varepsilon }(dS,T)  \\
&=\mathbb{E}[\left( S_{T+\varepsilon }-K\right) ^{+}] \\
&=C(K,T+\varepsilon ) .
\end{align*}%
Since, by assumption, $C$ is
continuous, convergence as $\varepsilon
\rightarrow 0$ to $C\left(
K,T\right) $ is trivial. 
\end{proof}

\begin{remark}
Instead of $\tau _{\varepsilon }:T\mapsto T+\varepsilon $ one may take any
strictly increasing $\tau _{\varepsilon }  :{ [0,\infty )}\rightarrow \lbrack
\varepsilon ,\infty )$; in general, by the chain rule from calculus, an additional factor $\tau _{\varepsilon
}^{\prime }$ will appear in Dupire's formula.
\end{remark}


\section{Appendix: Local volatility blowup in some jump models}

Throughout this section, we normalize spot w.l.o.g.\
to $S_0=1$.
At the money, i.e.\ for $K=S_0=1$, no blowup of the local volatility is to be expected.
For instance, in L\'evy jump diffusion models, $\sigma_{\mathrm{loc}}$
tends to the volatility~$\sigma$ of the
jump diffusion part as $T\to0$. To see this, recall the forward PIDE for the call price~\cite{BeCo10,CaGeMaYo04,CoVo05}:
\[
  \partial_T C = \frac12 K^2\sigma^2 \partial_{KK}C
  + \int_{-\infty}^\infty \nu(dz) \left( C(Ke^{-z},T) - C(K,T) - K(e^z-1)\partial_{K}C \right),
\]
where $\nu$ is the L\'evy measure. As $T\to0$, the integral tends to a non-negative constant.
The claim thus follows from~\eqref{DupForm}, since the density $\partial_{KK}C$ tends
to infinity for $K=S_0$.

By similar reasoning, we can quantify the off-the-money blowup of local vol, as soon as small-time asymptotics for the
density in the denominator of Dupire's formula~\eqref{DupForm} are available. For example, the density of
the NIG (normal inverse Gaussian) model is
$\sim const \cdot T$ (in fact, there is an explicit expression for the density~\cite{CoTa04}).
This implies $\sigma _{\mathrm{loc}}^{2}( K,T) \approx const/T$ for $K\neq S_0$. (Strictly speaking, the argument
gives only a \emph{bound} $O(1/T)$. To show that the numerator $\partial_T C$ tends to a
\emph{nonzero} constant seems difficult.)

We now obtain further asymptotic results by using the local volatility approximation
\begin{equation}\label{eq:wing}
\sigma _{\mathrm{loc}}^{2}( K,T) \approx \left. \frac{2\frac{\partial }{%
\partial T}m( s,T) }{s( s-1) }\right\vert _{s=\hat{s}%
( K,T) }
\end{equation}
 presented in~\cite{DeFrGe13,FrGe11}.
Here, $m(s,T)=\log M(s,T)$ denotes the log of the moment generating function
$M(s,T)=E[\exp(sX_T)]$ of the log-price $X_T=S_T$, and $\hat{s}=\hat{s}(K,T)$
solves the saddle point equation
\begin{equation*}
  \frac{\partial }{\partial s}m( s,T) = k := \log K.
\end{equation*}
While our focus in~\cite{DeFrGe13} was on asymptotics in~$K$, \eqref{eq:wing}
may as well be used for time asymptotics, provided that the underlying
saddle point approximation can be justified. The latter can be achieved by
analysing the Fourier representation of local vol, which is at the base of~\eqref{eq:wing}:
\begin{equation}\label{eq:fourier}
  \sigma_{\mathrm{loc}}^2(K,T) = \frac{2\partial_T C}{ K^2 \partial_{KK}C}
    = \frac{2 \int_{-i\infty}^{i\infty} \frac{\partial_T m(s,T)}{s(s-1)} e^{-ks}M(s,T) ds}
    {\int_{-i\infty}^{i\infty} e^{-ks} M(s,T) ds}.
\end{equation}
For instance, in~\cite{FrGe11} we discussed the strike asymptotics for Kou's
double exponential jump diffusion, and small-time asymptotics can be done
in the same way, the result being that local variance
is of order~$T^{-1/2}$.
Note that tail integral estimates are required to give a rigorous proof;
see~\cite{FrGe11,FrGeGuSt11} for two examples of such arguments. For the Kou
model, the moment generating function
  \[
    M(s,T) = \exp\left(T\left(\frac{\sigma^2 s^2}{2} + \lambda\left(
      \frac{\lambda_+ p}{\lambda_+ - s} +
      \frac{\lambda_-(1-p)}{\lambda_- + s}\right)\right)\right)
  \]
 has a singularity of the type ``exponential
of a pole'', and one can copy almost verbatim the saddle point analysis of~\cite{Ge11b}.
On the other hand, the NIG model, which we discussed above, is \emph{not} in the scope of~\eqref{eq:wing}.
 The reason
is simply that there is no saddle point, as the moment generating function
\begin{equation*}\label{eq:nig}
    M(s,T) = \exp\left( \delta T\left( \sqrt{\alpha^2-\beta^2}
      -\sqrt{\alpha^2-(\beta+s)^2} \right) \right)
  \end{equation*}
has no blowup at its singularity $\alpha-\beta$.

\subsection*{The Merton model}
The Merton jump
diffusion is another case in point for the saddle point approximation~\eqref{eq:wing}.
The  resulting off-the-money blowup is essentially of order $1/T$: 
\begin{equation}  \label{eq:merton blowup2}
\sigma_{\mathrm{loc}}^2(K,T) \sim const\cdot  \frac{|k|}{T} \left(\log\frac{|k|}{T}%
\right)^{-3/2}, \quad T\to0,
\end{equation}
where $k=\log K\neq0$ and the constant depends on the sign of~$k$. To see
this, recall
that the Merton jump diffusion has moment generating function 
\begin{equation*}
M(s,T)=\exp(T(\tfrac12 \sigma^2s^2+bs 
+\lambda(e^{\delta^2s^2/2+\mu s}-1)) 
\end{equation*}
with diffusion volatility~$\sigma$, jump intensity~$\lambda$, mean jump size~%
$\mu$ and jump size variance~$\delta^2$. The saddle point $\hat{s}$ solves $%
\partial_s m(s,1)=k/T$. As observed in~\cite{FrGe11}, for L\'evy models the saddle point
(if it exists) is always a function of $k/T$, so that various asymptotic regimes can
be captured by the same formula.
For $k>0$, the saddle point satisfies 
\begin{equation*}
\frac{\delta^2 \hat{s}^2}{2} = \log \frac{k}{T} - \frac{\sqrt{2}\mu}{\delta}
\sqrt{\log \frac{k}{T}} - \frac12 \log \log \frac{k}{T}  + \frac{\mu^2}{%
\delta^2}-\log\frac{\sqrt{2}}{\delta}  +O\left(\frac{\log \log \frac{k}{T}}{%
\sqrt{\log \frac{k}{T}}}\right) 
\end{equation*}
as $k/T\to\infty$. Inserting this into~\eqref{eq:wing} yields~%
\eqref{eq:merton blowup2}, with a similar reasoning for $k<0$.
As regards the tail integral estimates needed to make this rigorous, no unpleasant
surprises are to be expected, as double exponential singularities (here: at infinity)
are well known to be amenable to the saddle point method.

\subsection*{Merton's jump-to-ruin model}

In this model, the underlying $(S_t)$ follows Black-Scholes dynamics, but may
jump to zero (and stay there) at an independent exponential time
with parameter~$\lambda$. In other words, $\lambda$ is the risk-neutral
arrival rate of default. Out of the money, the additional default feature
has little influence on local vol for small~$T$, as it does not matter much that the
underlying can jump even further out of the money. So we expect
$\sigma_{\mathrm{loc}}^2(K,T) \to \sigma^2$ as $T\to 0$ for $K>S_0=1$.
Indeed, this follows immediately from the explicit formula (see~\cite{Ga06,La09})
\begin{equation}\label{eq:merton ruin}
  \sigma_{\mathrm{loc}}^2(K,T) = \sigma^2 +2\lambda\sigma \sqrt{T}
    \frac{N(d_2)}{N'(d_2)},
\end{equation}
where~$N$ is the standard Gaussian cdf and
\[
  d_2 = \frac{\log(S_0/K+\lambda T)}{\sigma \sqrt{T}} + \frac{\sigma \sqrt{T}}{2}.
\]
Formula~\eqref{eq:merton ruin} is an easy consequence
of the fact that the call price in this model
is just the Black-Scholes price with interest rate~$\lambda$ ($\lambda+r$ in general, but recall that we assume $r=0$ throughout).
The out-of-the-money convergence
$\sigma_{\mathrm{loc}}^2(K,T) \to \sigma^2$ can also be confirmed by our formula~\eqref{eq:wing}:
The moment generating function equals
\[
  M(s,T) =
  \begin{cases}
    \exp\left( T \left( \tfrac12\sigma^2s^2+(\lambda-\tfrac12 \sigma^2)s
    -\lambda \right)\right) & s\geq 0 \\
    \infty & s < 0.
  \end{cases}
\]
There is a saddle point $\hat{s}= k/(\sigma^2 T) + \tfrac12 - \lambda/\sigma^2$,
and hence~\eqref{eq:wing} gives $\sigma_{\mathrm{loc}}^2(K,T)=\sigma^2+O(T)$.

For $K<S_0=1$, on the other hand, \eqref{eq:merton ruin} reveals
a fast blowup of the order $\sigma_{\mathrm{loc}}^2(K,T) \approx e^{1/T}$
as $T\to0$. Formula~\eqref{eq:wing} seems not to be useful here,
as the moment generating function is not defined for $\mathrm{Re}(s)<0$,
since~$S_T$ may assume the value zero.

\bigskip

\textbf{Acknowledgment} PKF acknowledges support from MATHEON. We thank Peter
Laurence for sending us the unpublished preprint~\cite{La09}.

\bibliographystyle{siam}
\bibliography{../gerhold}











\end{document}